\documentclass{jfm}
\usepackage{graphicx}
\usepackage{epstopdf, epsfig}
\usepackage{amssymb}
\usepackage{amsmath}
\usepackage[dvipsnames]{xcolor}
\usepackage[]{lineno}
\usepackage[skip=0pt]{caption}

\usepackage{natbib}

\shorttitle{Effects of surface roughness on the propulsive performance  of pitching foils}
\shortauthor{Rodrigo Vilumbrales-Garcia et al.}

\title{Effects of surface roughness on the propulsive performance of pitching foils}

\author{Rodrigo Vilumbrales-Garcia\aff{1}\corresp{\email{r.vilumbrales-garcia@soton.ac.uk}},
  Melike Kurt\aff{1}, \newline Gabriel D. Weymouth\aff{1,2},
 \and Bharathram Ganapathisubramani\aff{1}}

\affiliation{\aff{1}Faculty of Engineering and Physical Sciences, University of Southampton, UK,\aff{2} Faculty of Mechanical, Maritime and Materials Engineering (3mE), TU Delft, NL}

\begin{document}
\maketitle

\begin{abstract}
The hydrodynamic influence of surface texture on static surfaces ranges from large drag penalties (roughness) to potential performance benefits (\textit{shark-like skin}). Although, it is of wide-ranging research interest, the impact of roughness on flapping systems has received limited attention. In this work, we explore the effect of roughness on unsteady performance of a harmonically pitching foil through experiments using foils with different surface roughness, at a fixed Strouhal number and within the Reynolds number ($Re$) range of $15k-30k$. The foils' surface roughness is altered by changing the distribution of spherical-cap shaped elements over the propulsor area. We find that the addition of surface roughness does not improve the performance compared to a smooth surface over the $Re$ range considered. The analysis of the flow fields shows near identical wakes regardless of the foil's surface roughness. The performance reduction mainly occurs due to an increase in profile drag. However, we find that the drag penalty due to roughness is reduced from $76\%$ for a static foil to $16\%$ for a flapping foil at the same mean angle of attack, with the strongest decrease measured at the highest $Re$. Our findings highlight that the effect of roughness on dynamic systems is very different than that on static systems, thereby, cannot be accounted for by only using information obtained from static cases. This also indicates that the performance of unsteady, flapping systems is more robust to the changes in surface roughness.
\end{abstract}

\section{Introduction} 
Surface roughness is ever-present in engineering applications leveraging fluid-structure interactions. Its implications on the flow and the consequent drag generation have been widely studied in the related literature. From the influence of roughness in pipe flow \citep{achenbach1971influence}, to its effects on the trajectory of a golf ball \citep{chowdhury2016study}, roughness plays a vital role in any application involving fluid-structure interaction considerations. For example, surface roughness can be detrimental to the performance of wind turbines. \cite{sagol2013issues} found that the accumulation of contamination agents in the blades leads to a reduction in power extraction, while, \cite{ehrmann2017effect} reported a performance decrease linked to an increase in roughness density and height. On the other hand, the use of roughness elements can lead to a drag reduction and certain performance gains for unsteady propulsion systems. Previous studies inspired from swimmers and flyers show that, from shark skin or dolphin skin \citep{dean2010shark, wainwright2019smooth} to feathers on a wing of a gliding bird \citep{van2015feather}, roughness in varying shapes and texture modifies the fluid flow over propulsor surfaces, leading to a reduction in drag or a decrease in flow separation. In engineering applications, \cite{gad1991separation} analysed the effects of roughness turbulators and found a $C_L/C_D$ increase when compared to a smooth foil for $Re\leq100000$. Also, the use of surface riblets can lead to a decrease in skin friction when aligned in the flow direction \citep{bechert1997experiments}, achieving a drag reduction of up to $8\%$ \citep{walsh1982turbulent}. When configured properly, surface roughness can be beneficial. It can reduce drag production and potentially improve the overall performance. Surface roughness can also have detrimental effects. Tailoring the surface roughness to have an improved performance requires a better understanding of the effect of shape, size, and area distribution of roughness elements on both the force production and the flow. 

The drag-reduction potential of surface roughness on aquatic swimmers have been explored mainly for static surfaces. For example, sharks can reduce their skin friction when their riblets are aligned with the flow \citep{dean2010shark}. \cite{bixler2013fluid} pointed out that the riblets lift and pin the vortices generated in the viscous sublayer, leading to a decrease in drag. \cite{bechert2000experiments} observed a drag reduction for interlocking 3D riblets.  \cite{afroz2016experimental} concluded that 'shark-like' textures can act like a passive flow separation control mechanism. \cite{du2022control} found smaller separated regions and adverse pressure gradients for the flow over a foil covered with tilted biomimetic shark scales. The effect of the shape and size of the rough elements were analysed by \cite{domel2018hydrodynamic}, highlighting the importance of the denticle shape, as they found a drag reduction only for the smaller of the three considered. Although surface roughness has shown promising potential for static bodies, its role in unsteady systems is still not clear. Shark-skin surfaces have been shown to increase the self-propelled swimming speed and reduce the drag of a flapping foil \citep{oeffner2012hydrodynamic, domel2018hydrodynamic}, but only when small denticles are used, whereas the larger elements can lead to an increase in drag. \cite{wen2014biomimetic} reported a reduction in energy consumption due to a formation of stronger leading-edge vortices. \cite{guo2021influence} found that, for static foils towed at a constant velocity, the roughness elements resulted in a considerably thicker boundary layer when compared to the smooth foil, while, for static foils in acceleration, the changes due to roughness in the wake characteristics were considerably smaller. Mostly, previous work conclude that shark-inspired surfaces can improve the performance of an unsteady body, but the potential benefit is strongly dependent on the shape and size of shark denticles, which often appear in highly complex geometries. Therefore, it is still to be seen if such performance improvement can be achieved with simple, commercially available roughness elements, located on the surface of an unsteady foil in harmonic motions.

In this study, we analyse experimentally the effects of surface roughness on the propulsive performance of a pitching foil by using simple roughness elements. In Section 2, we define the methodology and experimental setup used to actuate three different foils with varying roughness characteristics. We investigate the effects of Reynolds number in the range of $15,000 \leq Re \leq 30,000$ and report the propulsive performance of a pitching aerofoil in terms of thrust production ($C_X$) and efficiency ($\eta$). In Section 3, we detail the force and flow measurement results obtained for flapping foils, and draw a comparison between dynamic and static foil cases. 
\section{Experimental setup and methodology}\label{sec:methodology}
\begin{figure}
  \centerline{\includegraphics[width=0.9\textwidth]{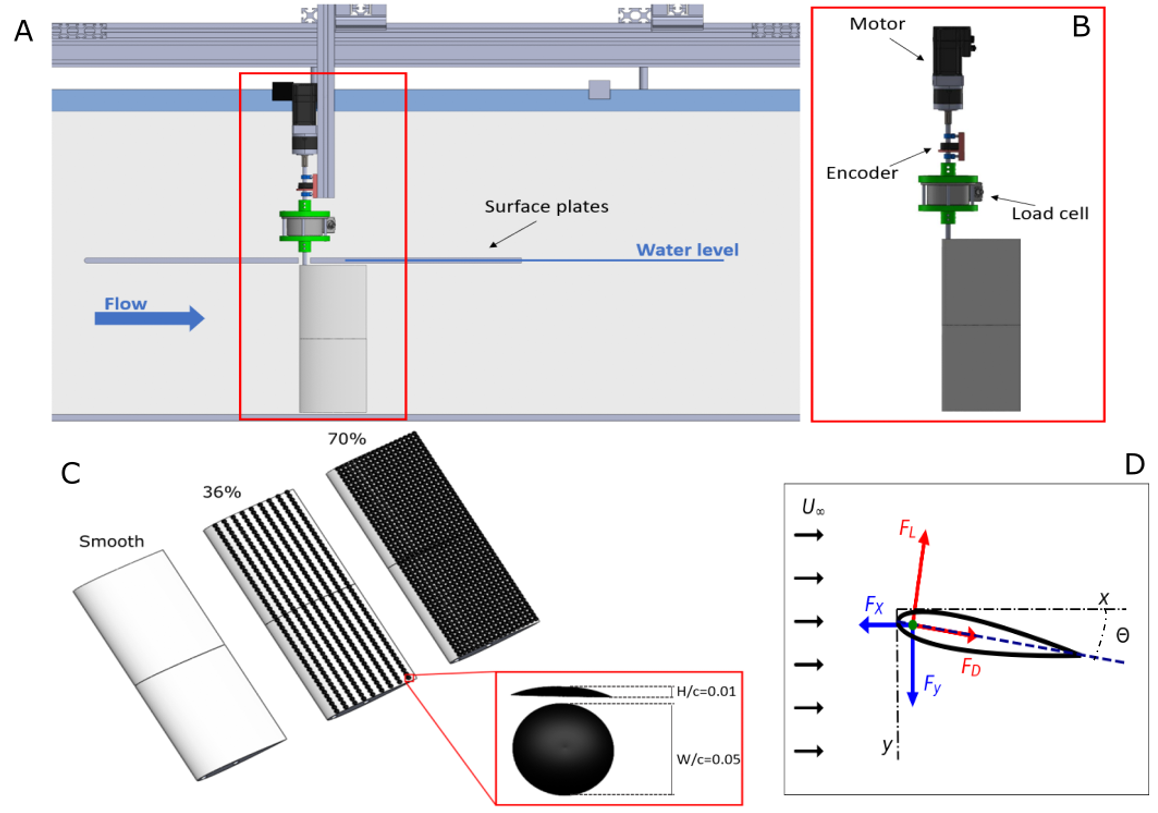}}
  \caption{Schematics of the experimental setup in the water flume (A), the actuation arm (B),  foils with three different roughness area coverage ratio (C), and the forces acting on the foil (D).}
\label{fig:setup}
\end{figure}
Force and flow measurements are conducted in a recirculating water flume at the University of Southampton, with a test section of 8.1 m length, 1.2m width and 0.9m depth. A surface plate is installed at the foil tip and the foil is placed right above the bottom wall to prevent tip vortex formation and enforce nominally two-dimensional flow over the foil as shown in Figure ~\ref{fig:setup}A.

Three foils with a rectangular planform and a NACA0012 cross-section were 3d-printed, with a chord-length of $c = 0.16m$ and an aspect ratio of $AR=2.5$. Spherical-cap shaped roughness element with a width (diameter) of $W = 0.05c$ and height of $H=0.01c$ \citep{dean2010shark} were placed on pressure and suction sides of the foils. As shown in Figure~\ref{fig:setup}C, in addition to the smooth foil, two different roughness levels are considered by varying the area occupied by the spherical-cap elements to $36\%$ and $70\%$ of the foil planform area.

Each foil was actuated with a stepper motor (Applied Motions STM23S) in sinusoidal pitching motions, about a point $0.08c$ distance from the leading edge. The prescribed motion is defined by  $\theta(t)=\theta_0 sin(2\pi f_0 t)$, where $\theta_0$ is the pitching amplitude and $f_0$ is the flapping frequency. The pitching amplitude $\theta_0$, Strouhal number $St=2Af_0 /U$, and reduced frequency $k=2\pi f_0 c/U$ were fixed all throughout the experiments at $\theta_0=7.5^\circ$, $St = 0.25$ and $k=6$, respectively, to ensure the foils to perform in the high-efficiency, thrust producing regime \citep{zurman2021fin, muscutt2017performance, kurt2018flow}. A Reynolds number sweep ($Re=Uc/\nu$ where $\nu$ is the kinematic viscosity) was conducted within the range of $15,000 \leq Re \leq 30,000$ by varying the flow velocity. In this range, the propulsive performance was previously reported to be $Re$ independent \citep{senturk2019reynolds}. A summary of the parameters used in this study is given in Table ~\ref{tab:matrix}.
\begin{table}
  \begin{center}
\def~{\hphantom{0}}
  \begin{tabular}{ccccc}
      $Re$  & 15,000   &  20,000 & 25,000 & 30,000\\
      $U$ [m/s]   & 0.10 & 0.14 & 0.17 & 0.21\\
      $f_0$ [Hz]   & 0.62 & 0.83 & 1.03 & 1.24\\
      $St$   & 0.25 & 0.25 & 0.25 & 0.25\\     
  \end{tabular}
  \caption{Experimental parameters used in the current study}
  \label{tab:matrix}
  \end{center}
\end{table}
The forces and moments acting on the foils were measured with a six-axis force sensor (ATI Gamma IP65). The motion was tracked using a rotary, incremental encoder (US Digital E5) attached on the motor shaft (Figure ~\ref{fig:setup} B). Each trial was conducted for a total of 100 flapping cycles and repeated five times. The measured forces were filtered using a Butterworth filter with a low-pass frequency of five times the flapping frequency. The power was calculated as a multiplication of pitching moment and the angular velocity which was derived from the measured angular displacement. The instantaneous and time-averaged performance metrics are the average values from 500 flapping cycles, measured over five trials. To distinguish instantaneous forces from time-averaged results, the latter is denoted by $\overline{(.)}$. The reported streamwise force (thrust) ($C_X$) and power ($C_P$) coefficients, and efficiency ($\eta$) are defined as,
 
 \begin{equation}
C_X = \frac{F_X}{\frac{1}{2}\rho U^2 c}, \qquad C_P = \frac{P}{\frac{1}{2}\rho U^3 c}, \qquad \eta=\frac{C_X}{C_P}
\end{equation}
\noindent where $\rho$ is the density of water and $U$ represents the free-stream flow velocity.

The force measurements were synchronised with planar Particle Image Velocimetry (PIV) measurements (cameras: {\it LaVision MX 4MP}, lasers: {\it Litron Nano PIV}). The field-of-view captures the entire foil and up to one chord-length in foil's wake. The software {\it Davis 10} was used to cross-correlate the acquired particle image pairs (with 24$\times$24 pixels with 50\% overlap). The flapping cycle was divided into twenty-two phases and twenty-five cycles were acquired per phase. The velocity fields corresponding to each phase was then averaged over 25 cycles.  
\section{Results}

\subsection{Flow-field and force production analysis of foils with different roughness area coverage ratios}

\begin{figure}
  \centerline{\includegraphics[width=0.9\textwidth]{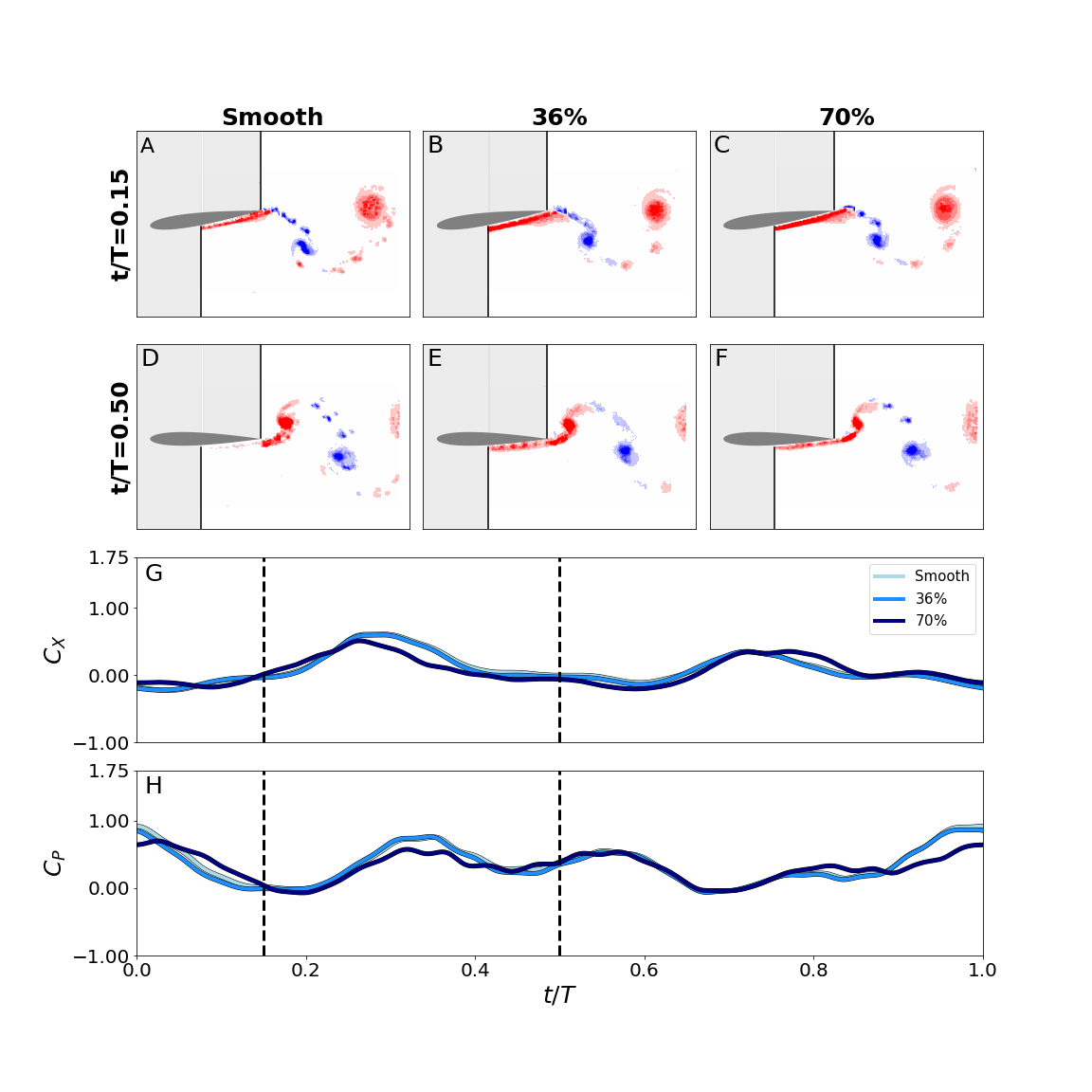}}
  \vspace*{-10mm}
  \caption{PIV results for $Re=25000$. $t/T=0.15$ (A,D,G) and $t/T=0.50$ (B,E,H) for the Smooth (A,D), $36\%$ (B,E) and $70\%$ (C,F). Instantaneous $C_X$ (H) and instantaneous $C_P$ (I)}
\label{fig:PIVhighRe}
\vspace{-5mm}
\end{figure}

Figure~\ref{fig:PIVhighRe} compares the out-of-plane vorticity and the instantaneous performance coefficients, $C_X$ and $C_P$ for all the roughness cases considered at $Re=25.000$. The first column (A,D), the second column (B,E) and the third column (C,F) present the evolution of the vorticity field around three pitching foils with $0\%$, $36\%$ and $70\%$ surface roughness at $t/T=0.15$, and $t/T=0.50$, respectively. Surprisingly, change in the roughness does not lead to any significant alteration in the vorticity fields. Regardless of the roughness coverage, all foils produce a reverse von Karman-street where two counter-rotating vortices per flapping cycle are shed from the trailing edge into the wake, as widely observed in the related literature for smooth foils \citep{muscutt2017performance, kurt2018flow, zurman2021fin}. Figure~\ref{fig:PIVhighRe} G-H presents the evolution of cycle-averaged thrust and power coefficients over one flapping cycle. Similar to the flow fields, the performance coefficients show only minor differences between the smooth foil and the foils with roughness. Although, we have only revealed the analysis associated with a single $Re$, these results hold across the $Re$ range considered here. In the supplementary material, we present the evolution of the flow-field over one flapping cycle at $Re=15,000$ and $Re=25,000$ as videos for comparison. Overall, these results from force and flow field measurements show that incorporating surface roughness does not have a strong influence on the development of the wake. Other parameters, such as Strouhal number or kinematics \citep{schnipper2009vortex} are known to significantly affect the evolution of the vortex structures, which can minimise the adverse effects on performance induced by the roughness elements.

\begin{figure}
  \centerline{\includegraphics[width=0.8\textwidth]{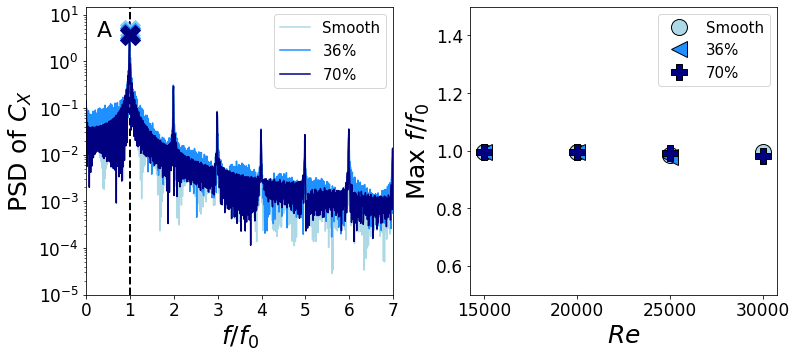}}
  \caption{(A) Fast Fourier Transform (FFT) analysis of the instantaneous $C_X$ at $Re=25,000$. The cross indicates the location of the peak for each case. The vertical dashed bar denotes $f/f_0=1$. (B) Peak frequency across the $Re$ values considered. A value of 1 denotes that the thrust force signal peak $f$ is equal to the input pitching frequency $f_0$.}
\label{fig:spectral}
\end{figure}

Figure~\ref{fig:spectral} introduces the spectral analysis conducted for $C_X$ presented in Figure~\ref{fig:PIVhighRe} G-H. The power spectra of the thrust force at $Re=25,000$ is shown in (A). The crosses indicate the location of the peak frequency for each foil. In (B), we introduce the dominant frequency ratio in the form of $f/f_0$, where $f_0$ is the prescribed pitching frequency across the $Re$ range considered. This analysis shows that the dominant frequency in thrust production corresponds to the pitching frequency $f_0$ for all the $Re$ values, and contains similar energy density for all the foils. This result combined with the similarities observed in both the wake and the instantaneous forces indicates that the performance of the foils is highly dominated by $f_0$, hence, by the kinematics. The dominant effects of the frequency and the kinematics observed in our study are similar to the findings by \cite{zurman2021fin}, who reported that compared to flapping frequency and kinematics, shape-related parameters such as sweep angle, have negligible effects on propulsive performance.

\begin{figure}
  \centerline{\includegraphics[width=1\textwidth]{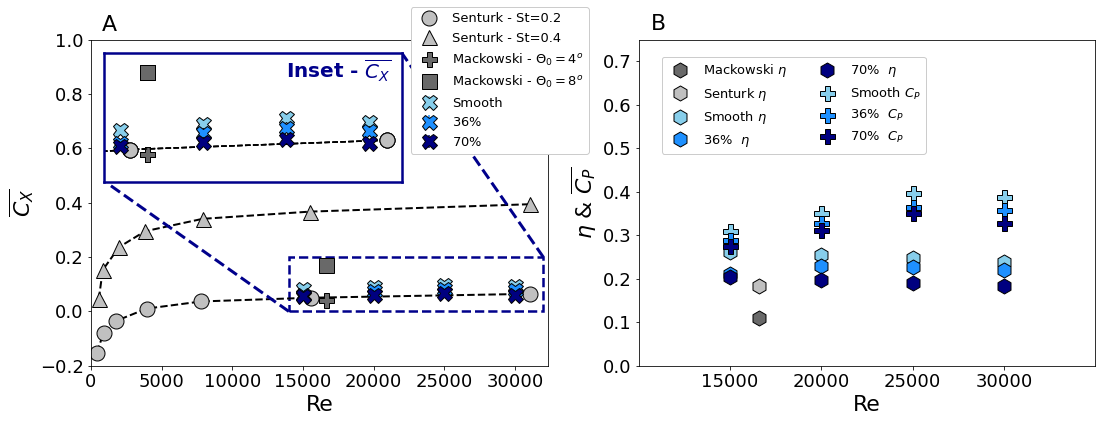}}
  \caption{A) $\overline{C_X}$ obtained in the current study (blue range) and compared with previous studies against $Re$: \cite{senturk2019reynolds} (gray) and \cite{mackowski2015direct} (dark-gray). The data enclosed by the blue box presents an inset of $\overline{C_X}$ data for the $Re$ range of $15,000\leq Re\leq30,000$. B) $\overline{C_P}$ results (hexagon) and $\eta$ (cross) for current and previous studies, marked by the same colours used in A.}
\label{fig:cyclemeans}
\end{figure}

Figure~\ref{fig:cyclemeans} presents the change in cycle-averaged performance coefficients for foils with different roughness coverages against $Re$. Starting with $\overline{C_X}$ (Figure~\ref{fig:cyclemeans}A), we compare our results with other NACA0012 studies conducted by \cite{mackowski2015direct} ($Re=16,600$, $0.1\leq St\leq0.4$), and \cite{senturk2019reynolds} ($500\leq Re\leq36,000$, $0.2\leq St\leq0.6$). Thrust, $\overline{C_X}$, obtained for the smooth foil increases slightly with Reynolds number, a trend similar to previous studies. The thrust values obtained at $St=0.25$ in the current study fall within the findings by \cite{senturk2019reynolds} at $St=0.2$ and $St=0.4$. In the inset enclosed by a blue box, it is shown that $C_X$ decreases with the addition of surface roughness across the $Re$ range. In Figure~\ref{fig:cyclemeans}B, our results indicate higher efficiency values than \cite{mackowski2015direct} and \cite{senturk2019reynolds}, which could be due to differences in the pivot point location (at $0.08c$ distance from the leading edge here, and at $0.25c$ distance in previous studies). For each roughness case, $\overline{C_P}$ slightly increases against $Re$, but regardless of the $Re$, increase in roughness causes a decrease in $\overline{C_P}$. The efficiency also decreases as the surface roughness increases, similar to thrust and power. Although the flow fields show negligible alterations with the change in roughness, the cycle averaged forces point to a performance reduction as the roughness increases. The thrust decrease observed for $36\%$ and $70\%$ roughness coverages compared to smooth foil can be related to an increase in the profile drag. To further explore this effect, in the next figure, we have compared our flapping foil results with static foil measurements carried out using the same foils within the same $Re$ range.

\subsection{Comparison between static and flapping regimes}
\begin{figure}
  \centerline{\includegraphics[width=0.75\textwidth]{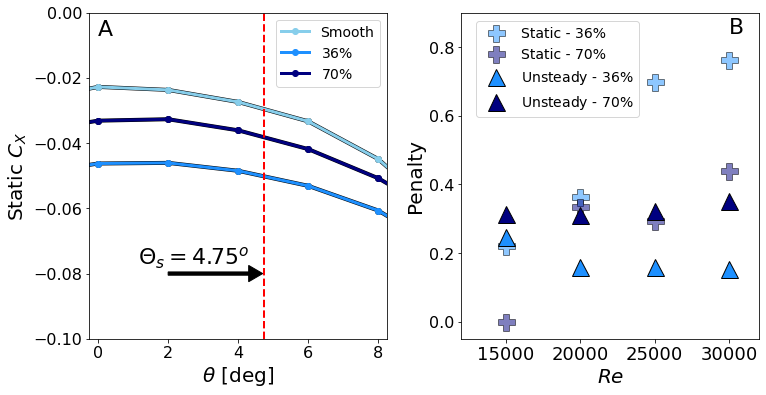}}
  \caption{A) static $C_X$ vs angle of attack $\Theta$ measured using static foils at $Re=25,000$. Smooth foil presented in light-blue, $36\%$ in medium-blue, and $70\%$ in dark-blue. The red dashed-line indicates the $\theta$ used to compare with the unsteady regime, defined as $\theta_s = 4.75^\circ$ B) Thrust and drag penalty due to roughness for both the flapping (triangles) and static results (crosses) for the $36\%$ case (medium blue) and the $70\%$ case (dark blue).\}}
\label{fig:pseudosteady}
\end{figure}

In this section, we introduce the data collected for static foils and compare it with the pitching foil results to further explore why there is a change in thrust production with a change in roughness coverage. The static data was acquired within the same $Re$ range and roughness coverages as the flapping cases, for an angle of attack ($\theta$) range of $-4^\circ\leq\theta\leq 20^\circ$. To compare both scenarios, we have selected an angle of attack value equal to the average $\theta$ experienced by the foil during half the pitching cycle (red dashed-line in Figure~\ref{fig:pseudosteady}A, denoted as $\theta_{s}$). Next, we develop a comparison parameter or \textit{penalty}, that evaluates the change in streamwise force generated by the smooth and rough foils. Given that static state will produce drag (for all three foils) and the flapping cases produce thrust, we present the penalty in its absolute value to help with the comparison. Since we have found surface roughness to be detrimental for $\overline{C_X}$ for all cases considered, a positive penalty value in the static state indicates an increase in drag due to roughness elements, while $Penalty>0$ in the flapping regime means a decrease in thrust caused by the roughness elements. Here, $Penalty$ is defined as the relative change in thrust for a rough foil compared to the smooth, $|(C_{X,rough}-C_{X,smooth})|/C_{X,smooth}$.

The penalty parameter is presented in Figure~\ref{fig:pseudosteady} for static foil (crosses) and flapping foil measurements (triangles). The addition of surface roughness increases the drag production in the static state across the $Re$ range considered. At $Re=30,000$, it reaches a $76\%$ drag penalty for the $36\%$ roughness and $43\%$ penalty for the $70\%$ roughness coverage, compared to the smooth foil. On the other hand, the flapping foils with roughness generate less thrust across the $Re$ range compared to the smooth foil. At $Re=30,000$, the thrust decreases by $35\%$ and $16\%$ for $70\%$ and $36\%$ roughness coverages, respectively. However, the flapping state appears to be more robust to $Re$ changes. It reduces the penalty observed for static foils, especially for the $36\%$ coverage. At $Re=30,000$, foils with $36\%$ coverage experiences a roughness penalty of $76\%$ in the static state compared to a $16\%$ penalty in the flapping state, which could be explained by the dominant effect that $St$ and the kinematics have on force production.

\begin{figure}
  \centerline{\includegraphics[width=1\textwidth]{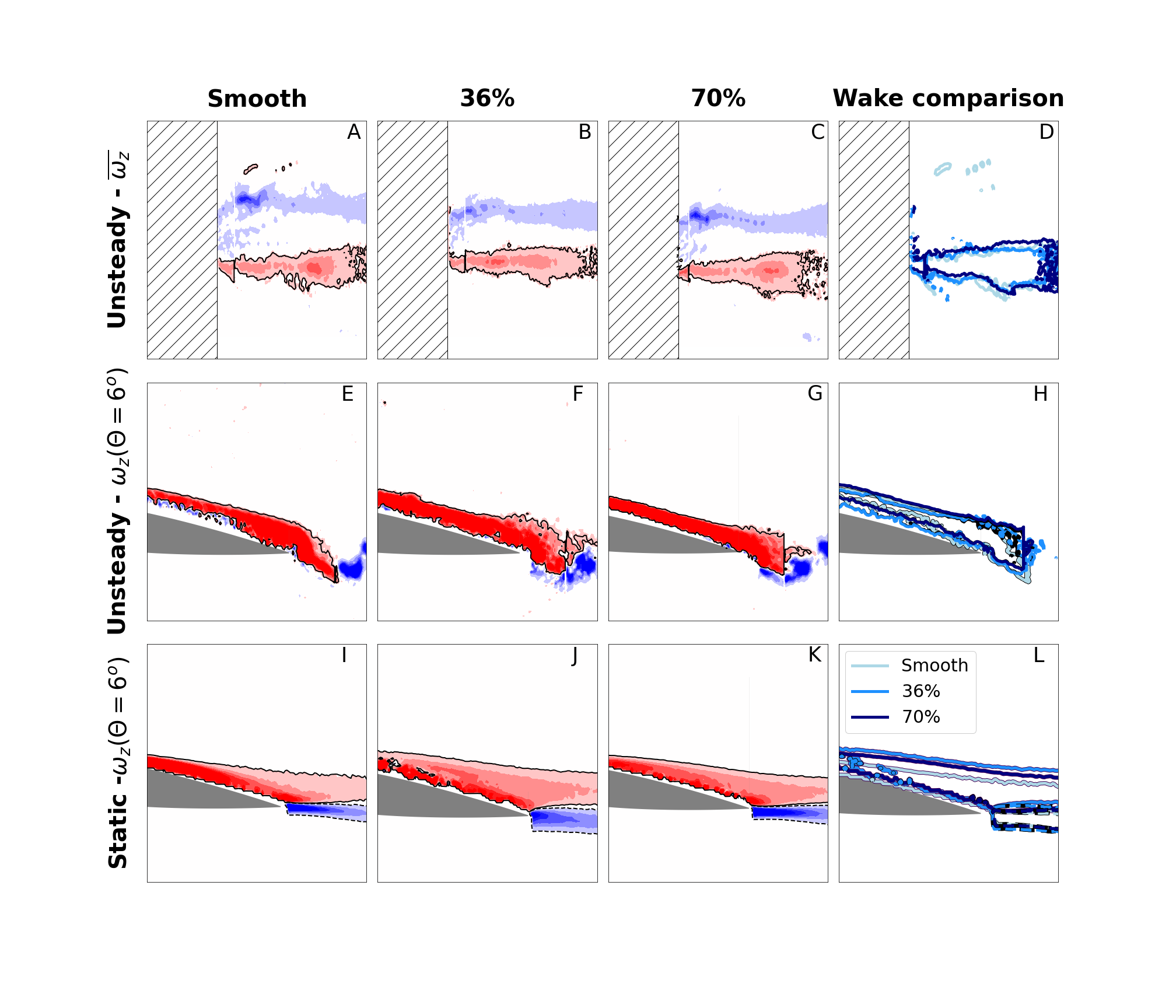}}
    \vspace{-10mm}
  \caption{Pitching cycle-averaged vorticity (A,B,C), flapping instantaneous vorticity at $\theta = 6^\circ$ degrees (E,F,G) and static PIV results at $\theta = 6^\circ$ (I,J,K). Smooth foil (A,E,I), $36\%$ (B,F,J), and $70\%$ (C,G,K). The comparison of the wakes generated by the foils at each of the conditions is presented at (D,H,L) }
\label{fig:BLdetail}
\end{figure}

To further analyse the data presented in Figure~\ref{fig:pseudosteady}, we present the out-of-plane vorticity ($\omega_Z$) in Figure~\ref{fig:BLdetail}. The first row consists of the cycle-averaged unsteady pitching $\omega_Z$, and the positive vorticity regions are enclosed with isolines. In second and third rows, we present the flow-field data measured at $\alpha=6^\circ$ for a pitching foil and a static foil, respectively. The first three columns correspond to $0\%$ (smooth), $36\%$ and $70\%$ roughness coverages, respectively. The fourth column introduces a comparison between different roughness cases with overlapped $\omega_Z$ isolines. The comparison of all three flapping cases suggests that the addition of surface roughness does not introduce major changes in shedding shear layers. In contrast, in the static state, foils with $36\%$ and $70\%$ roughness have thicker shear layer in time-average compared to the smooth case, similar to the findings by \cite{guo2021influence}. The presence of thicker shear layers for the roughness cases can be the culprit of $76\%$ drag penalty shown in Figure \ref{fig:pseudosteady}.
\vspace{-0.1in}
\section{Conclusions} \label{conclusions}
In this study, we have analysed the influence of surface roughness on the propulsive performance of flapping foils, using force and flow measurements. 
Three NACA0012 foils with different roughness coverage ratios have been constructed, and tested within the Reynolds number range of $15,000\leq Re \leq30,000$. 
We have found that the addition of surface roughness is detrimental to thrust production and efficiency of a pitching foil. The foils with $36\%$ and $70\%$ roughness produce $16\%$ and $35\%$ less thrust, respectively, compared to the smooth foil. We have determined that $Re$ does not play an important role on neither the thrust nor efficiency for the $Re$ range and roughness coverage ratios considered. Although we have seen no significant change in the wake flow, the foils with roughness experience a decrease in thrust and efficiency, which can be explained by an increase in profile drag associated with the roughness elements.
We have compared the effects of roughness on static and flapping states, finding that the former is considerably more sensitive to it. The roughness penalty for $36\%$ roughness coverage is reduced from $76\%$ in the static state to $16\%$ for flapping. The strongest decrease occurs at the highest Re, highlighting that the effect of roughness on flapping systems is very different than on static systems. This shows that the performance of flapping systems is more robust to the changes in surface roughness. \\
\noindent
\textbf{Declaration of Interests}: The authors report no conflict of interest.
\noindent
\textbf{Acknowledgements} \\
This research was supported financially by the Office of Naval Research Global Award N62909-18-1-2091, the Engineering and Physical Sciences Research Council (Grant No: EP/R034370/1) and the doctoral training award. \\

\noindent
\textbf{Data availability statement} \\
All data supporting this study will be made openly available from the University of Southampton repository upon publication.

\bibliographystyle{jfm}
\bibliography{jfm}

\end{document}